\documentclass[aps]{revtex4}%
\usepackage{amsfonts}
\usepackage{amsmath}
\usepackage{amssymb}
\usepackage{graphicx}%
\begin{document}
\title[Theory of turbulence]{Self-consistent theory of turbulence}
\author{S.A. Ktitorov}
\affiliation{A.F. Ioffe Physico-Technical Institute of the Russian Academy of Sciences,
Polytechnicheskaja str. 26, St. Petersburg, 194021, Russia}

\begin{abstract}
A new approach to the stochastic theory of turbulence is suggested. The
coloured noise that is present in the stochastic Navier-Stokes equation is
generated from the delta-correlated noise allowing us to avoid the nonlocal
field theory as it is the case in the conventional theory. A feed-back
mechanism is introduced in order to control the noise intensity.

\end{abstract}
\maketitle





\section{Introduction}

The modern theory of turbulence is based mainly on the stochastic processes
theory \cite{verma}. One of the constructive approaches to the developed
turbulence is based on a formal likeness between this problem and the dynamics
of phase transitions \cite{vas} (we cite here mostly this enciclopedean book
instead of a lot of original works critically reviewed there). The basic
equation of the phase transition dynamics reads:%

\begin{equation}
\partial_{t}\phi\left(  x\right)  =-\alpha\frac{\partial U\left(
x,\phi\right)  }{\partial\phi}+\eta\left(  x\right)  , \label{dyn}%
\end{equation}
where $\phi$ is the fluctuating order parameter, $\alpha$ is the Onsager
kinetic coefficient; $x\equiv\left(  \mathbf{r,}t\right)  $; $U\left(
x,\phi\right)  $ is a functional of $\phi$; in the simple case of the Brownian
particle it is reduced to a potential energy of the particle; $\eta\left(
x\right)  $ is a random force. It is usually assumed to be the Gaussian
stochastic process with the correlator%

\begin{equation}
\left\langle \eta\left(  x\right)  \eta\left(  x^{\prime}\right)
\right\rangle =2\alpha T\delta\left(  x-x^{\prime}\right)  , \label{delta}%
\end{equation}
where $T$ is the absolute temperature; $\delta\left(  x\right)  =\delta\left(
\mathbf{r}\right)  \delta\left(  t\right)  $, the transverse projector
$P_{ij}=(\delta_{ij}-\frac{k_{i}k_{j}}{k^{2}})$ appears in the case of the
incompressible liquid that is assumed here. The factor $2\alpha T$ before the
delta function is a manifestation of the fluctuation-dissipation theorem
\cite{fdt}.

The Navier-Stokes equation was similarly transformed into the stochastic
differential equation \cite{vas} by adding the random forces $\xi_{i}$:%

\begin{equation}
\rho\nabla_{t}v_{i}\left(  x\right)  =\nu\nabla^{2}v_{i}\left(  x\right)
-\partial_{i}p\left(  x\right)  +\xi_{i}\left(  x\right)  , \label{navier}%
\end{equation}
where $\nabla_{t}\equiv\partial_{t}+(\mathbf{v\cdot\nabla)}$, $\rho$ is the
mass density, $v_{i}$ is the transversal vector field of the velocity;
$div\mathbf{v}\equiv0$; $\nu$ is the kinematic viscosity; $p\left(  x\right)
$ is the pressure. The noise $\xi_{i}$ does not satisfy here the
fluctuation-dissipation theorem and its correlator doesn not look like
(\ref{delta}). The following form of the correlator is usually assumed%

\begin{equation}
\left\langle \xi_{i}(\mathbf{r,}t)\xi_{j}(\mathbf{r}^{\prime},t^{\prime
})\right\rangle =P_{ij}D(\mathbf{r}-\mathbf{r}^{\prime})\delta(t-t^{\prime}),
\label{turbnoise}%
\end{equation}
\bigskip but zero correlation time does not seem to be necessary. The
transverse projector $P_{ij}=(\delta_{ij}-\frac{k_{i}k_{j}}{k^{2}})$ appears
in the case of the incompressible liquid that is assumed here. The spatial
correlator $D(\mathbf{r})$ is usually chosen in the form \cite{vas}%

\begin{equation}
D(\mathbf{r})=\int\frac{d^{d}k}{\left(  2\pi\right)  ^{d}}\exp\left(
i\mathbf{kr}\right)  N(\mathbf{k}), \label{fourier}%
\end{equation}
\bigskip where $d$ is the syetem dimension, $N(\mathbf{k})$ is the pumping
function. Basing on Kholmogorov's ideas, this function is usually chosen so
that the pumping is essentially infrared. This is the point where a similarity
between these two problems terminates. Two forms of the pumping function are
the most popular now in the stochastic theory of turbulence:%

\begin{equation}
N(\mathbf{k})=\frac{D_{0}k^{4-d}}{(k^{2}+m^{2})^{\epsilon}}
\label{pumpmassless}%
\end{equation}
and%

\begin{equation}
N(\mathbf{k})=D_{0}k^{4-d-2\epsilon}, \label{pumpmass}%
\end{equation}
where $\epsilon$ is a formal parameter; the theory is logarithic at
$\epsilon=0$.

\section{Equation for the noise}

Admitting all the arguing in favour of these forms we are ready now to answer
the following question: how the noise $\xi(\mathbf{r},t)$ and its correlator
$D(\mathbf{r},t)$ can be obtained at least at the phenomenological (with
respect to the Navier-Stokes equation) level? Our opinion is that there exists
such equation, which can transform the weak delta-correlated (white noise)
Langevin noise $\eta$ into the strong long-distant correlated noise $\xi$, and
what is more, all necessary ingredients for it are well known. We are not
ready to present a consequtive theoretical derivation of such equation now;
that is why we will write an equation that is likely to exist by our opinion
and then we shall give some plausible reasoning in favour of it. The required
equation reads%

\begin{equation}
\gamma\frac{d\xi_{i}}{dt}-\alpha\Delta\xi_{i}+\beta\left[  \mathbf{v}\right]
\xi_{i}+\delta\xi_{j}\xi_{j}\xi_{i}=\eta_{i} \label{general}%
\end{equation}
where $\beta\left[  \mathbf{v}\right]  $ is a functional of the velocity
distribution. It provides a feedback: the velocity fluctuation intensity
controls the noise generation process. A form of this functional is not known,
but we can assume the following local one:%

\begin{equation}
\beta\left[  \mathbf{v}\right]  =\beta^{\prime}\frac{R_{c}^{2}-R^{2}}{R^{c}}.
\label{average}%
\end{equation}

Here $R$ and $R_{c}$ are respectively the local and threshold Reynolds
numbers. At small Reynolds numbers we can neglect the nonlinear term; then we
obtain the Langevin-type equation (pumping equation) transforming the Gaussian
white noise $\mathbf{\eta}$ into the Gaussian coloured pumping noise
$\mathbf{\xi}$:%

\begin{equation}
\gamma\frac{d\xi_{i}}{dt}-\alpha\Delta\xi_{i}+\xi_{i}=\eta_{i}. \label{linear}%
\end{equation}
If $\alpha$ and $\gamma$ are small enough, the pumping noise $\mathbf{\xi}$
remains really colourless like the thermodynamic noise $\mathbf{\eta}$. Notice
that such equation can be a useful tool in order to avoid a non-locality of
the theory induced by the coloured noise in the standard model (\ref{navier})
substituting the weak thermodynamic noise $\mathbf{\eta}$ with the strong
delta-correlated one, which however does not satisfy the
fluctuation-dissipation relation. In order to obtain the pumping equation,
which admits a solution of the form (\ref{pumpmass}) or (\ref{pumpmassless}),
we must substitute the Laplacian in the pumping equation with the fractionary
one \cite{samko}, \cite{fracginz}. We must admit $\gamma=0$ in order to obtain
the noise with the delta time correlation as it is usually the case in the
turbulence theory. When $R^{2}$ approaches $R_{c}^{2}$, one may not neglect
colouring anyhow, and what is more important, nonlinear effects became
essential so that we must return to (\ref{general}). The pumping noise becomes
non-Gaussian and coloured. The nonlinearity of this kind was really introduced
for the first time by L.D. Landau \cite{Land} for the complex amplitudes near
the threshold in the form:%

\begin{equation}
\frac{d\left|  A\right|  ^{2}}{dt}=a\left|  A\right|  ^{2}-b\left|  A\right|
^{4}. \label{land}%
\end{equation}
This equation is obviously equivalent (for real $a$ and $b$) to the
time-dependent Ginzburg - Landau equation (without spatial derivatives):%

\begin{equation}
\frac{dA}{dt}=\frac{a}{2}A-\frac{b}{2}\left\vert A\right\vert ^{2}. \label{GL}%
\end{equation}
Our equation (\ref{general}) differs mainly by a presence of the random force
$\eta$ and of the Laplacian. However, this work of Landau is not the only one
that can be considered as a precursor of our approach. There is a paper by
D.N. Zubarev et al \cite{zub}, where the Ginzburg-Landau equation was
introduced for the Reynolds stress tensor. It is possible that there is a
close relation between that and our approaches, but our equations are more
useful for application of the renormalization group technique. The important
advantage of our approach is its locality. No given nonlocal correlator is
present in the theory. Then the equation for the noise can be written in the form%

\begin{equation}
-\alpha\Delta^{\epsilon}\xi_{i}+\beta\frac{v_{c}^{2}-v^{2}}{v^{c}}\xi
_{i}+\delta\xi_{j}\xi_{j}\xi_{i}=\eta_{i}, \label{basic2}%
\end{equation}
where $\Delta^{\epsilon}$ is the fractionary Laplacian \cite{fracginz}.A final
conclusion on the realistic form of equations can be done only deriving them
from the basic "microscopic" Navier-Stokes equation. The equation set
(\ref{navier}), or (\ref{basic2}) can be solved applying the well developed
renormalization group technique \cite{vas}. However, it is interesting to
consider qualtatively, how the white noise $\eta(x,t)$ is transformed by the
nonlinear system (\ref{general}). The equation (\ref{general}) is well known
in the stochasic resonance theory for distributed systems \cite{stochres}. In
the absence of a regular signal this equation describes random jumps from one
well to another via the potential barrier. Correlation functions for such
processes are well studied \cite{stochres}. Now we can write the stochastic
path integral and the corresponding effective action:%

\begin{equation}
Z=\int D\eta DDvD\psi_{1}D\psi_{2}\exp\left[  -S_{eff}\left\{  \eta,\xi
,v,\psi_{1},\psi_{2}\right\}  \right]  ,\label{path}%
\end{equation}

\begin{align*}
S_{eff}\left\{  \eta,\xi,v,\psi_{1},\psi_{2}\right\}   &  =\int d^{d}x\psi
_{1}\left[  \rho\nabla_{t}v_{i}\left(  x\right)  -\nu\nabla^{2}v_{i}\left(
x\right)  -\xi_{i}\left(  x\right)  \right]  +\\
&  +\int d^{d}x\psi_{2}\left[  -\alpha\Delta^{\epsilon}\xi_{i}+\beta
\frac{v_{c}^{2}-v^{2}}{v^{c}}\xi_{i}+\delta\xi_{j}\xi_{j}\xi_{i}-\eta
_{i}+\frac{1}{2}\eta_{i}(\mathbf{r})P_{ij}D\eta_{j}(\mathbf{r})\right]  ,
\end{align*}
where $D$ is a constant characterizing intensity of nonrenormalized
fluctuations, $\psi_{1}$and $\psi_{2}$ aare stochastic auxiliary fields.

In conclusion, we have proposed a novel model for description of the developed
turbulence, which does not contain a fixed coloured noise that makes the
theory more attractive due to its locality.

\end{document}